# Deep learning for subgrid-scale turbulence modeling in large-eddy simulations of the atmospheric boundary layer

Yu Cheng[1], Marco Giometto[2], Pit Kauffmann[3], Ling Lin[3], Chen Cao[3], Cody Zupnick[3], Harold Li[1], Qi Li[4], Ryan Abernathey[5], Pierre Gentine[1]

[1]Department of Earth and Environmental Engineering, Columbia University, New York, NY 10027

[2]Department of Civil Engineering and Engineering Mechanics, Columbia University, New York, NY 10027

[3]Capstone Project at Data Science Institute, Columbia University, New York, NY 10027

[4]School of Civil and Environmental Engineering, Cornell University, Ithaca, NY 14853

[5]Department of Earth and Environmental Sciences, Columbia University, Palisades, NY 10964

Corresponding author: Pierre Gentine (pg2328@columbia.edu)

**Abstract**

In large-eddy simulations, subgrid-scale (SGS) processes are parameterized as a function of filtered grid-scale variables. First-order, algebraic SGS models are based on the eddy-viscosity assumption, which does not always hold for turbulence. Here we apply supervised deep neural networks (DNNs) to learn SGS stresses from a set of neighboring coarse-grained velocity from direct numerical simulations (DNSs) of the atmospheric boundary layer at friction Reynolds numbers $Re_\tau$ up to 1243 without invoking the eddy-viscosity assumption. The DNN model was found to produce higher correlation of SGS stresses compared to the Smagorinsky model and the Smagorinsky-Bardina mixed model in the surface and mixed layers and can be applied to different grid resolutions and various stability conditions ranging from near neutral to very unstable. The additional information on potential temperature and pressure were found not to be useful for SGS modeling. Deep learning thus demonstrates great potential for LESs of geophysical turbulence.

# 1 Introduction

Turbulent flows are ubiquitous in nature and in engineering systems. Geophysical turbulence at very high Reynolds numbers ($Re \sim 10^7$ in the atmospheric boundary layer [*Bradley et al.*, 1981]) exhibits a large range of scales that cannot be resolved by direct numerical simulations (DNSs) of the Navier-Stokes equations with current computational resources [*Pope*, 2000; *Speziale*, 1991; *Voller and Porte-Agel*, 2002]. The large-eddy simulation (LES) technique [*Deardorff*, 1970; *Lilly*, 1967] only resolves large-scale, filtered, eddy-motions and parameterizes the impact of subgrid-scale (SGS) turbulence [*Smagorinsky*, 1963] as a function of the resolved flow. Therefore, unlike DNSs that resolve all scales of motions, LESs do not face similar computational limitation due to the Reynolds number but are highly dependent on the accuracy of SGS modeling.

Widely used SGS models, such as the dynamic Smagorinsky model [*Germano et al.*, 1991; *Lilly*, 1992], the Smagorinsky-Bardina mixed models [*Bardina et al.*, 1980] and Wall-Adapting Local Eddy-viscosity (WALE) models [*Nicoud et al.*, 1999], are first-order, algebraic closure models, and rely on the eddy-viscosity assumption (a gradient-diffusion model), which is justified only if the mixing length is much smaller than the scale of the mean flow variations. However, this condition is not always met [*Schmitt*, 2007], so that there is no clear length scale separation [*Bernard and Handler*, 1990; *Corrsin*, 1975; *Egolf*, 1994; *Lilly*, 1967; *Speziale*, 1991; *Tennekes et al.*, 1972] for turbulent flows. Previous studies [*Clark et al.*, 1979; *Liu et al.*, 1994] showed that the correlation $\rho$ of SGS stress calculated from the Smagorinsky model and from DNSs was only around 0.2. The Bardina similarity model [*Bardina et al.*, 1980] resulted in a correlation coefficient $\rho$=0.8, amongst the highest correlations obtained in *a priori* tests. The Smagorinsky-Bardina mixed model [*Bardina et al.*, 1980] was proposed shortly after to enhance the dissipative properties of the standard Bardina model, while still preserving a high correlation around 0.8 in neutrally-stratified, incompressible channel flow [*Horiuti*, 1989].

The atmospheric boundary layer (ABL) can be unstable (stable) with vertical motion strengthened (damped) when there is heating (cooling) from the surface. In LESs relying on scale-invariant Smagorinsky models, scales of motion smaller than the grid stencil [*Canuto and Cheng*, 1997; *Lilly*, 1967] should fall within the isotropic inertial subrange [*Kolmogorov*, 1941]. As the Dougherty-Ozmidov length scale [*Dougherty*, 1961; *Grachev et al.*, 2015; *Ozmidov*, 1965] is typically modeled rather than resolved due to the limitation of computational resources [*Cheng et al.*, 2018], LESs of the strongly stable ABL can hardly resolve isotropic turbulence and stably stratified conditions will not be considered in this manuscript. On the other hand, LESs have been widely applied to study the unstable ABL [*Deardorff*, 1972; *Li et al.*, 2018; *Moeng*, 1984; *Nieuwstadt et al.*, 1993] as the energy-containing eddies are dominant and larger compared to those of neutral and unstable ABLs. Yet, various unstable (from weakly to highly unstable) conditions can fundamentally alter turbulent transport and its coherent structures. The eddy-viscosity formulation in SGS models can be modified to account for such stability effects [*Chung and Matheou*, 2014; *Mason*, 1989; *Mason and Brown*, 1999], but the quality and reliability of available parameterizations remain an open problem [*Chamecki et al.*, 2007; *Chung and Matheou*, 2014; *Kleissl et al.*, 2003].

Recently, deep neural networks (DNNs) with convolutional operations [*Krizhevsky et al.*, 2012; *LeCun et al.*, 1989] have been successfully applied for recognition and detection of images and objects [*Girshick et al.*, 2014; *Sermanet et al.*, 2013; *Taigman et al.*, 2014]. In general DNNs with convolutional operators are efficient (as they reduce the dimensionality) of data with a known, grid-like topology [*Goodfellow et al.*, 2016]. Moreover, *Bar-Sinai et al.* [2018] showed that DNNs can represent spatial gradients well. Recent work have shown that DNNs may perform well in learning the ABL turbulence structures and how they relate to the flow gradients, which are key to SGS stresses. For instance, *Ling et al.* [2016] successfully applied DNNs to model Reynolds stress in Reynolds-averaged Navier-Stokes (RANS) models. *Yang et al.* [2019] developed artificial neutral networks for wall modeling in LESs for neutral channel flow. *Gamahara and Hattori* [2017] trained an artificial neural network to represent SGS stresses in LESs for a channel flow and the gradient was imposed and calculated rather than learned. Compared to gradient-based method, additional non-adjacent neighboring pixels can be added to include non-local effects (e.g. curvature) in DNNs.

Here we apply DNNs to learn the SGS stress in LESs (details in Supporting Information) with varying degrees of stability from near neutral to very unstable conditions ($z_i/L = -678.2$, where $z_i$ is boundary layer height and $L$ Obukhov length [*Obukhov*, 1946]) in the ABL, using a series of coarse-grained DNSs as the training dataset. The results of the DNN models are compared to those of the Smagorinsky model [*Lilly*, 1967; *Smagorinsky*, 1963] and of the Smagorinsky-Bardina mixed model [*Bardina et al.*, 1980]. One focus of this study is to gain insight into DNN-based SGS modeling across stability conditions and grid resolutions. We also examine if the DNN model could be applied to higher Reynolds numbers by evaluating the predicted SGS stresses in the log-law region [*von Kármán*, 1930] of wall turbulence (details in Supporting Information [*Monin and Obukhov*, 1954; *Obukhov*, 1946; *Panofsky*, 1963; *Paulson*, 1970]), which is "universal" [*Marusic et al.*, 2013] across high Reynolds numbers. On the one hand, *a posteriori* tests (SGS models applied to an LES experiment to calculate stresses) can hardly provide insights [*Meneveau and Katz*, 2000] into the detailed physics of SGS models due to combined effects of numerical discretization, time integration and average. On the other hand, *a priori* studies (SGS models applied to DNS data to calculate stresses) are a more fundamental prerequisite (compared to *a posteriori* tests) to understand the structure of the subgrid-scale model [*Meneveau and Katz*, 2000] and is of particular importance for stability-aware models [*Bou-Zeid et al.*, 2010; *Kleissl et al.*, 2004; *Porté-Agel et al.*, 2001; *Xu and Chen*, 2016] in LESs. We hence here limit our analysis to an *a priori* study, in line with previous efforts [*Clark et al.*, 1979; *Domaradzki et al.*, 1993; *Härtel et al.*, 1994; *McMillan and Ferziger*, 1979].

**2 Methodology**

2.1 Introduction to DNS data

Four different DNS datasets are processed to provide the input data for training and test, including three simulations of convective boundary layers and one simulation of neutral channel flow. Details of the DNS setups can be found in Table S1 and Supporting Information [*Chorin*, 1968; *Giometto et al.*, 2017; *Li et al.*, 2018; *Nieuwstadt et al.*, 1993; *Orlandi*, 2012; *Shah and Bou-Zeid*, 2014; *Wray*, 1990]. The viscous layer has been removed from the training and prediction datasets. In convective boundary layers, the initial velocity field is set by the geostrophic wind $U_g$. The Reynolds number is defined as $Re_\tau = \frac{u_\tau z_i}{\nu}$, where $u_\tau$ is the friction velocity, $z_i$ the boundary layer height and $\nu$ the kinematic viscosity. The three simulations of the convective boundary layers are named Sh2, Sh5 and Sh20, with corresponding shear Reynolds numbers $Re_\tau$ of 309 ($z_i/L = -678.2$), 554 ($z_i/L = -105.1$) and 1243 ($z_i/L = -7.14$), respectively. Each dataset consists of streamwise velocity $u$, spanwise velocity $v$, vertical velocity $w$, potential velocity $\theta$ and modified pressure $p^*$ fields at different time steps ($p^* = p + \frac{\rho}{2}u_l u_l$ [*Li et al.*, 2018]). For example, $Sh5_{t=1}$ denotes dataset Sh5 at time $t = 1$. $\Delta_z^+ \equiv \Delta_z u_\tau/\nu$ is the spatial grid resolution denoted in terms of inner units in the vertical direction. The fully-developed incompressible planar channel flow dataset [*Giometto et al.*, 2017] is named Channel1. The Reynolds number is defined as $Re_h = \frac{u_\tau h}{\nu} = 546$, where $h$ is the channel half-width.

Each DNS dataset is spatially filtered [*Leonard*, 1975] to provide coarse-grained data with different spatial grid resolutions in inner units. Similarly to *Clark et al.* [1979], we use a top-hat spatial filter [*Clark et al.*, 1979] (details in Supporting Information). The deviatoric part of SGS stress tensor $\tau_{lk}^d$ is parameterized in LESs [*Lilly*, 1967; *Smagorinsky*, 1963]. $\tau_{lk,DNS}^d$ is calculated using the DNS fine-grid velocity. The estimated stress by DNNs is named $\tau_{lk,NN}^d$. As an example, the velocity in the $x$-$y$ plane at $z^+ = \frac{z}{(\nu/u_\tau)} = 129.8$ in the logarithmic equilibrium layer of the original and spatially filtered dataset $Sh20_{t=1}$ are shown (Figure S1).

## 2.2 Deep neural networks

The DNNs are trained to reproduce the SGS stress $\tau_{lk,DNS}^d$ that are generated from the DNS data, i.e., we use supervised learning. For example, the SGS stress $\tau_{lk,DNS}^d$ at each spatial point is modelled using the coarse-grained variables $\overline{u_l}$, $\overline{v_l}$ and $\overline{w_l}$ ($\overline{\theta}$ or $\overline{p^*}$ is also added in some cases as the input of DNNs) in the neighboring $3 \times 3 \times 3$ box, which is used as the input layer of DNNs (schematic in Figure S2) with a total dimension of $3 \times 3 \times 3 \times 3$ (here the last "3" is number of input variables: $\overline{u_l}$, $\overline{v_l}$ and $\overline{w_l}$). Similarly to *Yang et al.* [2019], Galilean invariance and rotational invariance were not imposed, since an intrinsic velocity reference and a local coordinate system are given by the wall.

We tested different dimensions of the input box (Figure S3 and Supporting Information [*Goodfellow et al.*, 2016; *Hinton et al.*, 2012; *Ioffe and Szegedy*, 2015; *LeCun et al.*, 1989; *McCulloch and Pitts*, 1943; *Nair and Hinton*, 2010]). Increasing input dimension from $3 \times 3 \times 3$ to $5 \times 5 \times 5$ or $7 \times 7 \times 7$ does not lead to much improvement for the correlations of $\tau_{lk,DNS}^d$ and $\tau_{lk,NN}^d$. We therefore selected input dimension of $3 \times 3 \times 3$ for the final model as a tradeoff between high correlation of SGS stresses and requirements due to parallel computation of LESs. We have also tested several architectures (not shown here), including typical convolutional neural networks (CNNs) and other dimensions of input box but we finally chose the architecture with the highest correlation of $\tau_{lk,DNS}^d$ and $\tau_{lk,NN}^d$. We tested 8 different combinations of input datasets (with different stabilities), input variables ($u, v, w, \theta, p^*$) and dimensions of input box for training the DNNs (Table 1). The correlation coefficients [*Clark et al.*, 1979] of $\tau_{lk,DNS}^d$ and $\tau_{lk,NN}^d$ in the turbulent region and zooming into the log-law layer are used to evaluate the performance of the models.

**Table 1**. *Training datasets, input variables and prediction datasets of different DNN models.* Correlation coefficients of $\tau_{13,DNS}$ from DNS data and $\tau_{13,NN}$ from DNN models in the whole field of the coarse-grained DNS data are also shown. The first 6 DNN models have an input dimension of $3 \times 3 \times 3$, model M_uvw_multiSh_box555 has an input dimension of $5 \times 5 \times 5$ and model M_uvw_multiSh_box777 has an input dimension of $7 \times 7 \times 7$.

| DNN model name | Training dataset | Input variables | Prediction dataset | Correlation of $\tau_{13,DNS}$ and $\tau_{13,NN}$ |
|---|---|---|---|---|
| M_uvw_multiSh | $Sh20_{t=1,\Delta_z^+=21}$, $Sh20_{t=2,\Delta_z^+=21}$, $Sh2_{t=1,\Delta_z^+=6}$, $Sh2_{t=2,\Delta_z^+=6}$, | $u, v, w$ | $Sh20_{t=4,\Delta_z^+=11}$, $Sh20_{t=4,\Delta_z^+=21}$, $Sh20_{t=4,\Delta_z^+=42}$, $Sh20_{t=4,\Delta_z^+=64}$, $Sh2_{t=4,\Delta_z^+=6}$, $Sh5_{t=1,\Delta_z^+=10}$, $Channel1_{\Delta_z^+=23}$, | 0.834 0.840 0.816 0.778 0.824 0.818 0.415 |
| M_uvw | $Sh20_{t=1,\Delta_z^+=21}$ | $u, v, w$ | $Sh20_{t=4,\Delta_z^+=21}$, $Sh20_{t=4,\Delta_z^+=42}$, $Sh20_{t=4,\Delta_z^+=64}$, $Sh2_{t=4,\Delta_z^+=6}$, $Sh5_{t=1,\Delta_z^+=10}$, | 0.838 0.817 0.781 0.769 0.775 |
| M_uvw$\theta$ | $Sh20_{t=1,\Delta_z^+=21}$ | $u, v, w, \theta$ | $Sh20_{t=4,\Delta_z^+=21}$, $Sh20_{t=4,\Delta_z^+=42}$, $Sh20_{t=4,\Delta_z^+=64}$, $Sh2_{t=4,\Delta_z^+=6}$, $Sh5_{t=1,\Delta_z^+=10}$, | 0.843 0.825 0.798 0.777 0.787 |
| M_uvw$p$ | $Sh20_{t=1,\Delta_z^+=21}$ | $u, v, w, p^*$ | $Sh20_{t=4,\Delta_z^+=21}$, $Sh20_{t=4,\Delta_z^+=42}$, $Sh20_{t=4,\Delta_z^+=64}$, $Sh2_{t=4,\Delta_z^+=6}$, $Sh5_{t=1,\Delta_z^+=10}$, | 0.842 0.822 0.791 0.763 0.771 |
| M_uvw_multiT | $Sh20_{t=1,\Delta_z^+=21}$, $Sh20_{t=2,\Delta_z^+=21}$, $Sh20_{t=3,\Delta_z^+=21}$, $Sh20_{t=5,\Delta_z^+=21}$ | $u, v, w$ | $Sh20_{t=4,\Delta_z^+=21}$, $Sh20_{t=4,\Delta_z^+=42}$, $Sh20_{t=4,\Delta_z^+=64}$, $Sh2_{t=4,\Delta_z^+=6}$, $Sh5_{t=1,\Delta_z^+=10}$, | 0.850 0.829 0.797 0.814 0.787 |
| M_uvw_multiC | $Sh20_{t=1,\Delta_z^+=11}$, $Sh20_{t=1,\Delta_z^+=21}$, $Sh20_{t=1,\Delta_z^+=42}$, $Sh20_{t=1,\Delta_z^+=64}$, | $u, v, w$ | $Sh20_{t=4,\Delta_z^+=21}$, $Sh20_{t=4,\Delta_z^+=42}$, $Sh20_{t=4,\Delta_z^+=64}$, $Sh2_{t=4,\Delta_z^+=6}$, $Sh5_{t=1,\Delta_z^+=10}$, | 0.773 0.803 0.703 0.757 0.758 |
| M_uvw_multiSh_box555 | $Sh20_{t=1,\Delta_z^+=21}$, $Sh20_{t=2,\Delta_z^+=21}$, $Sh2_{t=1,\Delta_z^+=6}$, $Sh2_{t=2,\Delta_z^+=6}$, | $u, v, w$ | $Sh20_{t=4,\Delta_z^+=21}$, $Sh20_{t=4,\Delta_z^+=42}$, $Sh20_{t=4,\Delta_z^+=64}$, $Sh2_{t=4,\Delta_z^+=6}$, $Sh5_{t=1,\Delta_z^+=10}$, | 0.876 0.847 0.808 0.862 0.858 |
| M_uvw_multiSh_box777 | $Sh20_{t=1,\Delta_z^+=21}$, $Sh20_{t=2,\Delta_z^+=21}$, $Sh2_{t=1,\Delta_z^+=6}$, $Sh2_{t=2,\Delta_z^+=6}$, | $u, v, w$ | $Sh20_{t=4,\Delta_z^+=21}$, $Sh20_{t=4,\Delta_z^+=42}$, $Sh20_{t=4,\Delta_z^+=64}$, $Sh2_{t=4,\Delta_z^+=6}$, $Sh5_{t=1,\Delta_z^+=10}$, | 0.881 0.851 0.810 0.873 0.867 |

## 3 Results

3.1 DNNs trained on datasets with different stability conditions

The deviatoric SGS stresses of dataset $Sh5_{t=1,\Delta_z^+=10}$ calculated from the coarse-grained DNS data (denoted by $\tau_{lk,DNS}^d$), the Smagorinsky model [*Smagorinsky*, 1963] (denoted by $\tau_{lk,S}^d$), the Smagorinsky-Bardina mixed model [*Bardina et al.*, 1980] (denoted by $\tau_{lk,SB}^d$) and the DNN model M_uvw_multiSh (denoted by $\tau_{lk,NN}^d$) at $z^+ = \frac{z}{(\nu/u_\tau)} = 77.3$ in the log-law region are shown in Figure S4 for illustration purposes. The SGS stress $\tau_{13}$ averaged in the $x$-$y$ plane at different $z^+$ of dataset $Sh5_{t=1,\Delta_z^+=10}$ are compared (Figure 1a). The mean of $\tau_{13}$ is better captured by the DNN model in the majority of the domain including the log-law region between the two dotted lines, compared to the Smagorinsky model and the Smagorinsky-Bardina mixed model. The spatial variation of $\tau_{13}$ in the DNN prediction is similar to that in the DNS data, while the Smagorinsky model and the Smagorinsky-Bardina mixed model underestimate the variance of $\tau_{13}$ in the $z$ direction (Figure 1a) and do not capture the vertical trend. This is further confirmed by the probability distribution function (Figure 1b): the $\tau_{13}$ distribution in Smagorinsky and Smagorinsky-Bardina predictions are too concentrated around the mean compared to that in the DNS data and DNN prediction. The latter indeed exhibits a broader distribution and better captures the tails of the distribution (Figure 1b). The correlation of $\tau_{13,DNS}$ and $\tau_{13,S}$, $\tau_{13,SB}$ and $\tau_{13,NN}$ in $x$-$y$ plane at different $z^+$ are compared (Figure 1c). The correlation of $\tau_{13,S}$ and $\tau_{13,DNS}$ is about 0.2 in the middle of the vertical domain, in agreement with previous studies [Clark et al., 1979; Liu et al., 1994], while the correlation of $\tau_{13,NN}$ and $\tau_{13,DNS}$ is over 0.8, which is about 0.05 higher than that of the Smagorinsky-Bardina mixed model [Bardina et al., 1980; Horiuti, 1989] in the majority domain including the log-law region and the mixed layer that is above the log-law region.

The streamwise power spectra of $\tau_{13,DNS}$, $\tau_{13,S}$, $\tau_{13,SB}$ and $\tau_{13,NN}$ averaged in the $x$-$y$ plane at $z^+ = \frac{z}{(v/u_\tau)} = 77.3$ in the log-law region are then compared (Figure 1d). The power spectra produced by the DNN model are very close to the DNS data across the normalized wavenumber domain $0.05 < kz < 6$, where $k$ is the streamwise wavenumber and $z$ is distance to the wall. However, the power spectra produced by the Smagorinsky model and Smagorinsky-Bardina mixed model deviate much more from the DNS data compared to the DNN model, especially for $0.05 < kz < 0.3$ and $kz > 4$. Therefore, the DNN model M_uvw_multiSh predicts more accurately the SGS stress tensor components when compared to the Smagorinsky model and the Smagorinsky-Bardina mixed model both in the whole DNS field and the log-law region.

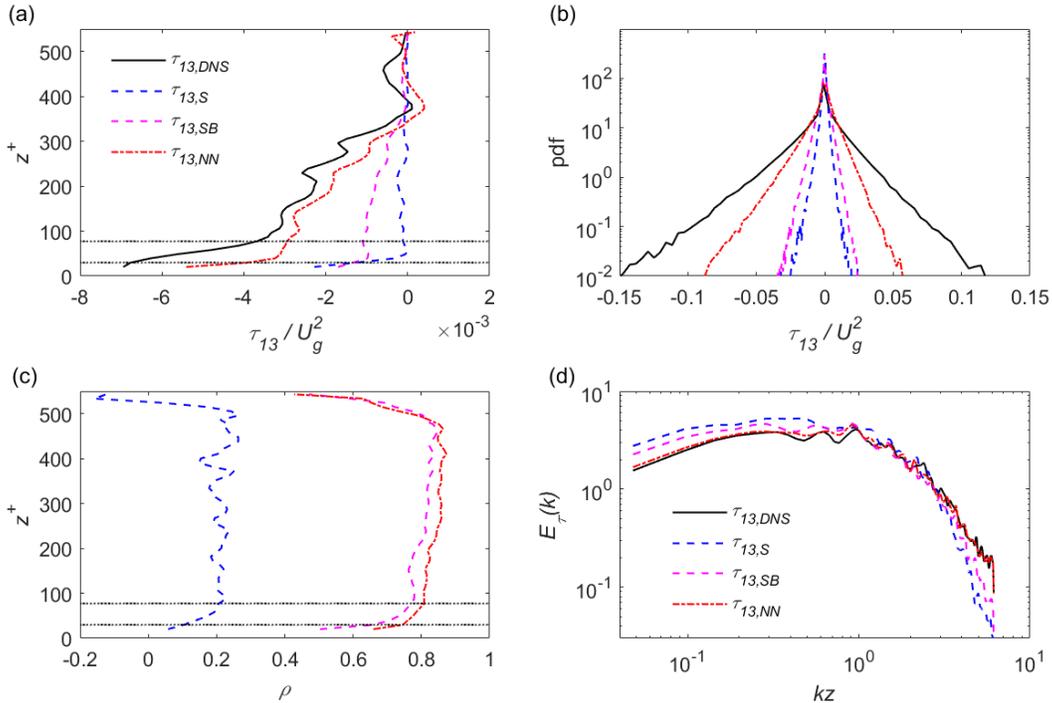

**Figure 1.** Comparison of $\tau_{13,DNS}$, $\tau_{13,S}$, $\tau_{13,SB}$ and $\tau_{13,NN}$. (a) Mean SGS stresses of $x$-$y$ plane at different $z^+$. (b) Probability distribution function (pdf) of SGS stresses in the whole DNS field.

(c) Correlation of SGS stresses in $x$-$y$ plane at different $z^+$. (d) Streamwise spectra of SGS stresses averaged in $x$-$y$ plane at $z^+ = \frac{z}{(\nu/u_\tau)} = 77.3$ in the log-law region. The dataset is $Sh5_{t=1,\Delta_z^+=10}$ and DNN model is M_uvw_multiSh. $U_g$ is geostrophic wind, $\rho$ is correlation, $\nu$ is kinematic viscosity, $u_\tau$ is friction velocity, $k$ is streamwise wavenumber, $z$ is distance to the wall and $E_\tau$ is power spectra. $z^+$ between the two dotted lines is the log-law region in (a) and (c).

The correlation coefficients $\rho$ calculated from $\tau_{13,S}$, $\tau_{13,SB}$ and $\tau_{13,NN}$ in the whole DNS field are compared (Figure 2a) across different test datasets. The overall correlation coefficients $\rho \approx 0.2$ between $\tau_{13,S}$ and $\tau_{13,DNS}$, while $\rho \approx 0.8$ between $\tau_{13,NN}$ and $\tau_{13,DNS}$ except for dataset $Channel1_{\Delta_z^+=23}$ which has a different boundary condition compared to other DNS datasets. The produced correlation of the DNN model is around 0.05 higher than that of the Smagorinsky-Bardina mixed model [*Bardina et al.*, 1980]. Similar results can be found for the correlation only for the log-law region (Figure S6), emphasizing the quality of the DNN across turbulent regions of the flow.

To further evaluate the potential of the DNN model, we focus on the correlation $\rho$ in the log-law region (Table S2). The DNN model M_uvw_multiSh produces smaller $\rho$ for $\tau_{13}$ (Table S2) for most prediction datasets in the log-law region compared to the whole DNS turbulent region. Model M_uvw_multiSh (trained on datasets with $\Delta_z^+ = 21$ and $\Delta_z^+ = 6$) produces a correlation of 0.780 in the log-law region for $Sh5_{t=1,\Delta_z^+=10}$, which is high considering previous studies [*Bardina et al.*, 1980; *Clark et al.*, 1979; *Horiuti*, 1989]. The largest correlation decrease in the log-law region compared to the whole DNS field is around 0.170 for the dataset $Sh20_{t=4,\Delta_z^+=64}$, where there is though only three vertical grid levels in the former. In fact, the model M_uvw_multiSh produces correlation coefficients not smaller than 0.780 in all datasets with more than 5 vertical grid levels in the log-law region. Therefore, we conclude that the DNN model M_uvw_multiSh can reproduce SGS stresses well in the log-law region.

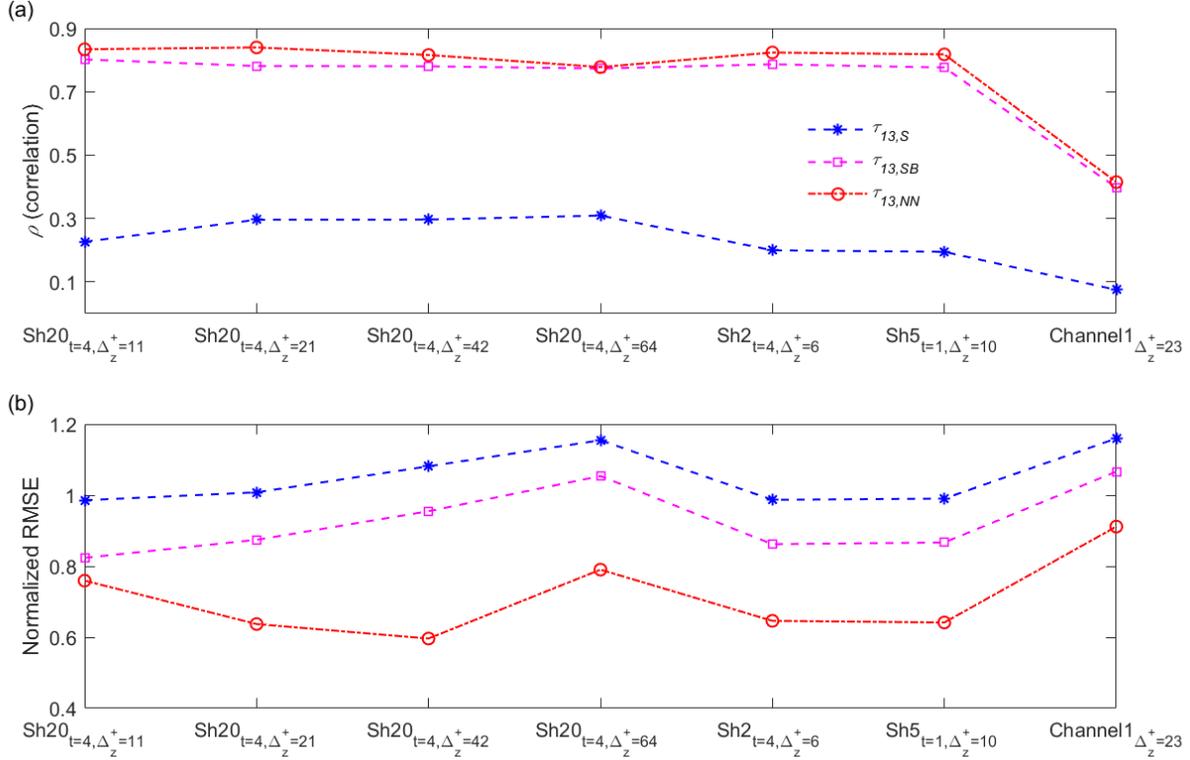

**Figure 2**. (a) Correlation coefficients calculated from $\tau_{13,S}$, $\tau_{13,SB}$ and $\tau_{13,NN}$ predicted by model M_uvw_multiSh for the whole DNS field. (b) Normalized root mean square error (RMSE) calculated from $\tau_{13,S}$, $\tau_{13,SB}$ and $\tau_{13,NN}$ for the whole DNS field. The prediction datasets are listed in $x$ axis.

The DNN model M_uvw_multiSh, the Smagorinsky model, and the Smagorinsky-Bardina mixed model all produce much higher root mean square error (RMSE) in the more buoyant cases: $Sh2_{t=4,\Delta_z^+=6}$ ($z_i/L = -678.2$), and $Sh5_{t=1,\Delta_z^+=10}$ ($z_i/L = -105.1$). The difference in RMSE across different stabilities is related to topological changes in the coherent structures in the unstable cases (transitioning from rolls to thermals) with highly varying $\tau_{lk,DNS}^d$ in space due to heating from the bottom. When RMSE is normalized by the standard deviation of $\tau_{lk,DNS}^d$ (we divide the RMSE by the standard deviation of $\tau_{lk,DNS}^d$ in the whole DNS turbulent region), its magnitude is of order 1 across all datasets (Figure 2b). Therefore, similarly to correlation coefficients that are normalized by the standard deviation, we also use normalized RMSE to evaluate the DNN performance and find that the DNN model produces the lowest normalized RMSE (Figure 2b).

The model M_uvw_multiSh is trained using both highly convective Sh2 ($z_i/L = -678.2$) and high-shear Sh20 ($z_i/L = -7.14$) but can correctly predict SGS stresses for Sh5 (moderately convective, $z_i/L = -105.1$) better than the Smagorinsky model and Smagorinsky-Bardina mixed model in terms of correlation, probability distribution, mean value, power spectra and normalized RMSE. Therefore, it appears that the DNN model M_uvw_multiSh has learned the topology of the turbulent structures from the two extreme shear and convective cases and can appropriately be used to make predictions on datasets with intermediate stability. In other words the model trained on extreme cases can interpolate turbulence behavior in between.

3.2 Sensitivity to input parameters

In addition to using $u$, $v$ and $w$ as the input to the DNNs, we also added potential temperature or modified pressure as an input variable for the training data $Sh20_{t=1,\Delta_z^+=21}$. The resulted correlation of models M_uvw$\theta$ and M_uvw$p$ (Table 1) with additional inputs are compared to that of the model trained only with $u$, $v$, $w$ (model M_uvw). As a reference, we also show results of model M_uvw_multiSh (Figures 3a, S7a and S8a).

For datasets $Sh20_{t=4,\Delta_z^+=21}$, $Sh20_{t=4,\Delta_z^+=42}$ and $Sh20_{t=4,\Delta_z^+=64}$, model M_uvw$\theta$ (using $u$, $v$, $w$ and $\theta$ as input) produces a slightly higher correlation for $\tau_{13}$ (Figure 3a) but slightly lower correlation for $\tau_{23}$ (Figure S8a) compared to model M_uvw (using $u$, $v$ and $w$ as input). For datasets $Sh2_{t=4,\Delta_z^+=6}$, and $Sh5_{t=1,\Delta_z^+=10}$, model M_uvw$\theta$ produces close correlation for $\tau_{13}$ and $\tau_{23}$ (Figures 3a and S8a) compared to model M_uvw. In the log-law region, models M_uvw$\theta$ and M_uvw produce very close correlation for $\tau_{13}$ (Table S2) with differences less than 0.02, so that the effects of potential temperature is not significant. Therefore, adding potential temperature as an extra input variable to include stability effects is not needed in such input boxes, across different unstable conditions. This is in contrast with some SGS models that additionally consider stability effects in unstable conditions [Mason, 1989; Mason and Brown, 1999] to calculate SGS stresses. Here the stability information is evidently already encoded in the velocity field.

For datasets $Sh20_{t=4,\Delta_z^+=21}$, $Sh20_{t=4,\Delta_z^+=42}$, and $Sh20_{t=4,\Delta_z^+=64}$, model M_uvw$p$ (using $u$, $v$, $w$ and $p^*$ as input) including pressure effects produces very close correlation for $\tau_{13}$ (Figure 3a), $\tau_{12}$ (Figure S7a) and $\tau_{23}$ (Figure S8a) compared to model M_uvw (using $u$, $v$ and $w$ as input). In the log-law region, models M_uvw$p$ and M_uvw produce very close correlation for $\tau_{13}$ (Table S2) with differences less than 0.02, so that the effects of pressure is not significant. *Bernard* *and Handler* [1990] suggested that pressure might contribute to acceleration transport, which could be a key component of the Reynolds stress. However, we here show that pressure need not be explicitly taken into account to calculate SGS stresses as the velocity field seems to have inherently incorporated pressure effects. Besides, potential temperature and pressure do not additionally increase the correlation by changing the input box to $5 \times 5 \times 5$ and $7 \times 7 \times 7$ (not shown here). Based on those results, we decided to only keep the velocity field as our main input.

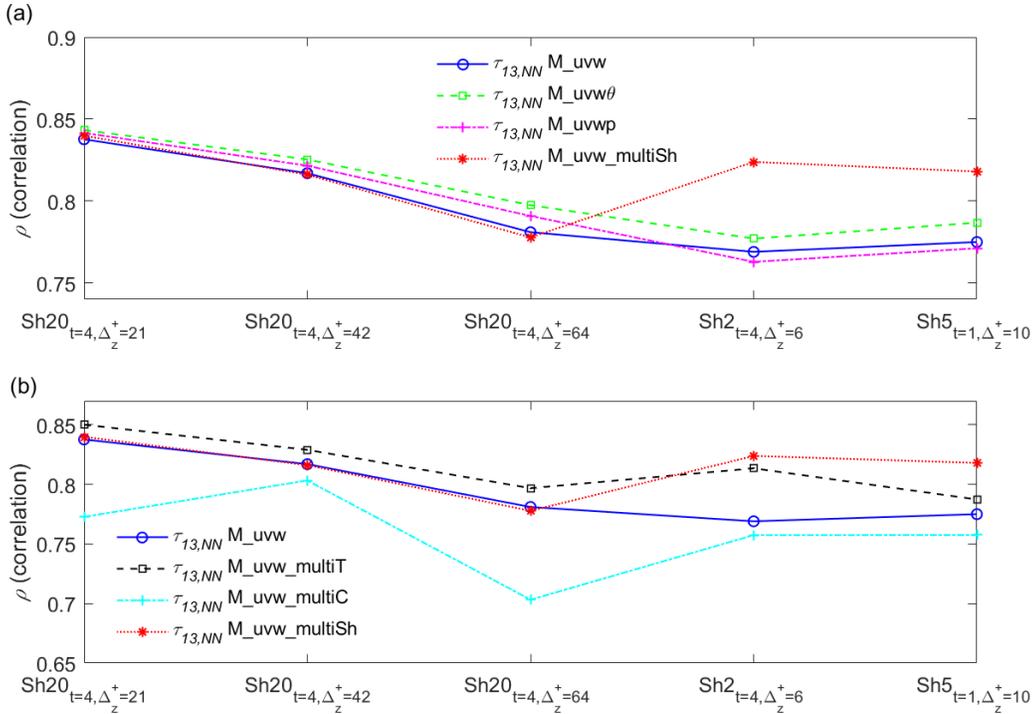

**Figure 3.** Correlation coefficients between $\tau_{13,DNS}$ and $\tau_{13,NN}$ predicted from different DNN models in the whole DNS field. (a) Comparison between the DNN models M_uvw_multiSh, M_uvw, M_uvw$\theta$ and M_uvw$p$. (b) Comparison between the DNN models M_uvw_multiSh, M_uvw, M_uvw_multiT and M_uvw_multiC. The prediction datasets are listed in $x$ axis.

3.3 Effects of the stability range of training data

To test if the initial training data need to cover different stability conditions to capture various coherent structures [*Agee et al.*, 1973; *Atkinson and Zhang*, 1996; *Li and Bou-Zeid*, 2011; *Li et al.*, 2018; *Lumley*, 1981; *Salesky et al.*, 2017], we also train a DNN model (denoted as model M_uvw_multiT) using only $Sh20_{\Delta_z^+=21}$ with high shear ($z_i/L = -7.14$) at different time steps.

For datasets $Sh20_{t=4,\Delta_z^+=21}$ ($z_i/L = -7.14$) and $Sh20_{t=4,\Delta_z^+=42}$ ($z_i/L = -7.14$), model M_uvw_multiT produces a higher correlation (around 0.02) for $\tau_{13}$ (Figure 3b) compared to model M_uvw_multiSh, since the former model only contains information of Sh20 ($z_i/L = -7.14$) while the latter model contains information from both Sh20 and Sh2 ($z_i/L = -678.2$). For dataset $Sh5_{t=1,\Delta_z^+=10}$ ($z_i/L = -105.1$), model M_uvw_multiT produces a lower correlation by around 0.04 for $\tau_{13}$ (Figures 3b) compared to model M_uvw_multiSh. The above analysis of the performance of models M_uvw_multiSh and M_uvw_multiT in in the whole DNS domain also holds for the log-law region (Table S2). Therefore, to make predictions on datasets with the same stability as existing training data, it is enough to just train on one dataset. However, to make predictions in other stability conditions, it is essential to train on a range of stability conditions from unstable to shear-driven conditions that cover the prediction stability.

3.4 Effects of grid resolutions – scale awareness

An important consideration for an SGS model is if it works well for LESs with different grid resolutions. Therefore, we also train model M_uvw_multiC on datasets with different grid resolutions. For the coarse-resolution dataset $Sh20_{t=4,\Delta_z^+=64}$ (Figure 3b and Table S2), model M_uvw (trained on $Sh20_{t=1,\Delta_z^+=21}$) produces a higher correlation by 0.078 for $\tau_{13}$ than model M_uvw_multiC (trained on $Sh20_{t=1,\Delta_z^+=11}$, $Sh20_{t=1,\Delta_z^+=21}$, $Sh20_{t=1,\Delta_z^+=42}$ and $Sh20_{t=1,\Delta_z^+=64}$) in the whole DNS field, although model M_uvw_multiC is trained on extra datasets with different grid resolutions. In particular, model M_uvw produces a higher correlation by 0.135 for $\tau_{13}$ compared to model M_uvw_multiC in the log-law region of dataset $Sh20_{t=4,\Delta_z^+=64}$. One possible explanation could be that the DNS datasets with coarser resolution (e.g., $Sh20_{t=1,\Delta_z^+=42}$ and $Sh20_{t=1,\Delta_z^+=64}$) lose some of the smaller-scale turbulence information thus leading to relatively worse performance when used to train DNN models.

For the fine-resolution dataset $Sh20_{t=4,\Delta_z^+=11}$ (Table S2), model M_uvw produces a lower correlation by 0.047 than model M_uvw_multiC, which is due to the fine-resolution training data $Sh20_{t=1,\Delta_z^+=11}$ applied to the latter. Including training data with the same grid resolution as test dataset can contribute to improved correlation for prediction on fine-resolution datasets. Therefore, training on datasets with the finest grid resolutions is able to produce good correlation of SGS stresses on datasets across different grid resolutions.

**4 Conclusions**

We use deep neural networks to construct SGS models for LESs of turbulence across stability conditions from near neutral to very unstable in the atmospheric boundary layer. The SGS stress at each spatial point is predicted using resolved velocity in the neighboring box in a similar way to convolutional operation. We also use the DNN model to test hypotheses on the structure and physics of the turbulence SGS model. Our primary objective is to shed light on the development of DNN-based SGS modeling across stability conditions in particular. The main findings and suggestions from our study are as follows.

1) SGS models based on DNNs produce more accurate SGS stresses compared to the Smagorinsky model and the Smagorinsky-Bardina mixed model in terms of correlation, normalized RMSE, mean value, probability distribution and power spectra. This was not necessarily expected *a priori*, given the fact that we used a regular mean square error cost function, which could for instance mostly target the mean state.
2) The DNN model trained on datasets with two extreme stability conditions (high shear and highly convective) performs well when applied to datasets with intermediate stability. This is of particular importance for stability-aware models in LESs.
3) The DNN model produces high correlation in the log-law region without inclusion of the distance to the wall.
4) The input dimension of $7 \times 7 \times 7$ and $5 \times 5 \times 5$ do not lead to much improvement over input dimension of $3 \times 3 \times 3$. This suggests that the SGS stress essentially depends on the local neighboring velocity field.
5) Potential temperature and modified pressure do not provide additional predictive skill beyond that provided by velocity, challenging the results of previous studies that additionally considered buoyancy (using potential temperature) effects in unstable conditions or that emphasized the importance of pressure on the Reynolds stress.
6) To make predictions on datasets with the same stability as existing training data, it is enough to just train on one dataset. To make predictions on other unstable cases at different stability conditions, it is better to train on a wide range of stability conditions that cover the prediction stability.

7) The DNN-based SGS model can be applied to a range of grid resolutions.

Our study suggests that DNN-based SGS models could eventually replace traditional SGS schemes in LESs of ABL turbulence, leading to better estimations of the SGS stress tensors. A key step towards this goal, which we will pursue in future works, is to assess the numerical stability and conservation properties of the DNN-SGS approach in an online simulation.

**Acknowledgments**

PG would like to acknowledge funding from the National Science Foundation (NSF CAREER, EAR-1552304) and from the Department of Energy (DOE Early Career, DE-SC00142013). We have no conflict to declare.

The code is accessible from the following link.

https://github.com/Codyz/dscaptstone/tree/master/Code/DNN/Final%20Models/Single%20Output

**Supporting Information**

This supporting information provides detailed introduction to SGS models for large-eddy simulations, setup of direct numerical simulations, the universal log-law region and setup of deep neural networks in this study.

**S1 Details of SGS LES model**

The spatially filtered Navier-Stokes equations in the Boussinesq approximation can be written as

$$\frac{\partial \overline{u_l}}{\partial t} + \frac{\partial (\overline{u_l}\,\overline{u_k})}{\partial x_k} = -\frac{1}{\rho_0}\frac{\partial \bar{p}}{\partial x_l} + \nu \frac{\partial}{\partial x_k}\left(\frac{\partial \overline{u_l}}{\partial x_k}\right) - \frac{\partial \tau_{lk}}{\partial x_k} + \delta_{l3} g \frac{\bar{\theta} - \bar{\theta}_{ref}}{\bar{\theta}_{ref}}, \qquad (1)$$

where $u$ is velocity, $x$ the Cartesian coordinate vector, $t$ time, $\rho_0$ the mean fluid density, $p$ pressure, $\nu$ the kinematic viscosity, $\tau$ the subgrid-scale stress, $l$ and $k$ indices, $\delta$ the Kronecker delta, $g$ gravity acceleration, $\theta$ fluctuating potential temperature, $\theta_{ref}$ mean potential temperature, and ($\overline{\cdots}$) a convolution filter [*Leonard*, 1975]. The SGS tensor $\tau_{lk}$ can be decomposed into a deviatoric component $\tau_{lk}^d$ and its trace $\tau_{mm}\frac{\delta_{lk}}{3}$,

$$\tau_{lk} \equiv \overline{(u_l u_k)} - \overline{u_l}\,\overline{u_k} = \tau_{lk}^d + \tau_{mm}\frac{\delta_{lk}}{3}, \qquad (2)$$

where $\tau_{mm}\frac{\delta_{lk}}{3}$ is typically absorbed into the pressure variable. $\tau_{lk}^d$ is parameterized in SGS models [*Lilly*, 1967; *Smagorinsky*, 1963] and is modelled by DNNs here.

**S2 Details of DNS setup**

Four different DNS datasets are processed to provide the input data for training and test, including 3 simulations of convective boundary layers ($z_i/L$ equals to $-7.14$, $-105.1$ and $-678.2$, respectively) and 1 simulation of neutral channel flow. The viscous layer has been removed from the training and prediction datasets. In the 3 datasets of boundary layer flow [*Li et al.*, 2018], the incompressible Navier-Stokes equations with Boussinesq approximation are solved. Details on the code can be found in *Shah and Bou-Zeid* [2014]. The boundary conditions used are: the bottom boundary is no-slip and no penetration while the top boundary layer is free-slip and no penetration. Similarly to *Li et al.* [2018], a sponge layer occupies the top 25% grid points in vertical direction to dissipate gravity waves [*Nieuwstadt et al.*, 1993]. The viscous and capping inversion layers are not used in the training or prediction datasets. Boundary conditions for the temperature field is constant flux at the surface and zero flux at the top of the computational domain. A mixed pseudospectral approach and second-order accurate staggered centered finite difference scheme are adopted to discretize the system of equations in the horizontal and vertical directions, respectively. A fractional-step solver is used to solve the system of equations and a second-order Adams-Bashforth scheme is used for time integration. The dataset Sh2 has grid points of $1200 \times 800 \times 602$ while both Sh5 and Sh20 have grid points of $nx \times ny \times nz = 1200 \times 800 \times 626$ in streamwise ($x$), spanwise ($y$) and vertical ($z$) directions, respectively. The choice of number of points in the vertical direction does not need to be constrained by consideration of Fast Fourier transform, which is only applied in the horizontal direction, since we use a second-order finite difference scheme.

The fully-developed incompressible planar channel flow dataset [*Giometto et al.*, 2017] is named Channel1, with grid points of $nx \times ny \times nz = 770 \times 770 \times 385$ in streamwise ($x$), spanwise ($y$) and vertical ($z$) directions. A fully staggered second-order centered finite difference method with grid stretching is used for the spatial discretization [*Orlandi*, 2012] and a third-order Runge-Kutta method [*Wray*, 1990] for the time integration. An operator splitting approach [*Chorin*, 1968] is also used to solve the system of equations. Only the bottom half grid-points are used as test datasets of DNNs since the flow is symmetric in the vertical direction. The lowest layer used for DNN prediction is $z^+ = \frac{z}{(\nu/u_\tau)} = 48.2$.

Similarly to [*Clark et al.*, 1979], the velocity $\bar{u}_l(x, y, z)$ with a coarse-graining factor of $2k + 1$ is calculated as

$$\bar{u}_l(x, y, z) = \frac{1}{(2k+1)^3} \sum_{x'=x-k}^{x'=x+k} \sum_{y'=y-k}^{y'=y+k} \sum_{z'=x-k}^{z'=z+k} u_l(x', y', z'), \qquad (3)$$

where $(x, y, z)$ are the coordinates of the points on the fine grid. The filter is spatial averaging of $2k + 1$ points in each direction. Unlike scale-invariant Smagorinsky models, the cutoff grid sizes of the DNN model need not fall in the inertial subrange. The cutoff grid sizes of coarse-grained datasets $Sh20_{t=4,\Delta_z^+=42}$ and $Sh20_{t=4,\Delta_z^+=64}$ are in the inertial subrange of the corresponding raw DNS data as in typical LESs, while the cutoff sizes of other coarse-grained DNSs are in the dissipation range (Table 1), which is due to the narrow inertial subrange of low DNS Reynolds numbers.

In the log-law region of wall turbulence, the mean velocity $U$ is related to the distance from the wall $z$ by a logarithmic function [*Marusic et al.*, 2013; *von Kármán*, 1930],

$$\frac{U}{u_\tau} = \frac{1}{\kappa}\log(z^+) + A, \qquad (4)$$

where $u_\tau$ is the friction velocity, $\kappa$ is the von Kármán constant, $z^+ = \frac{zu_\tau}{\nu}$, $z$ is the distance to the wall, $\nu$ is the kinematic viscosity and $A$ is a parameter that depends on the roughness of the surface. *Marusic et al.* [2013] showed the existence of a universal log-law region from boundary layers, pipe flow and the ABL at high Reynolds numbers, we thus assume that the DNN model could be applied to higher Reynolds numbers if it can capture the SGS stress well in this universal region. The log-law region of neutral channel flow can be found by plotting $\frac{U}{u_\tau}$ against $\frac{1}{\kappa}\log(z^+)$. For unstable boundary layers, the above relation has to be revised due to buoyancy fluxes from the bottom. According to Monin-Obukhov similarity theory [*Monin and Obukhov*, 1954], *Panofsky* [1963] obtained

$$U = \frac{u_\tau}{\kappa}\left(\log\left(\frac{z}{z_0}\right) - \psi_m\right), \tag{5}$$

where $z_0$ is the roughness length and $\psi_m$ satisfies [*Paulson*, 1970]

$$\psi_m = \log\left(\frac{1+x}{2}\right)^2 + \log\left(\frac{1+x^2}{2}\right) - 2\tan^{-1} x + \frac{\pi}{2}, \tag{6}$$

where $x = \left(1 - 16\frac{z}{L}\right)^{1/4}$ and $L$ is the Obukhov length [*Obukhov*, 1946]. According to the above equations, we plotted $\frac{\kappa U}{u_\tau} + \psi_m$ against $\log(z)$ to find the log-law region in convective DNS.

**S3 Details of DNN setup**

A mapping $y = f(x)$ is called network if $f(x)$ consists of a chain of activation functions, $f(x) = f^n\left(\cdots f^k\left(\cdots f^2(f^1(x))\right)\right)$, where $f^1, f^2, \ldots f^k, \ldots f^n$ are different activation functions [*Goodfellow et al.*, 2016] with $f^k$ the kth layer and $f^n$ the output layer. If the information only flows from $x$ to $y$ the networks are called feedforward [*Goodfellow et al.*, 2016]. The name neural networks arises because of the analogy with neural sciences [*McCulloch and Pitts*, 1943]. Deep neural networks (DNN) typically refer to neural networks with more than two layers [*Goodfellow et al.*, 2016].

The Python library Keras [*Chollet and Others*, 2015] with the Tensorflow [*Abadi et al.*, 2016] backend has been used for the DNN setup. The first dense layer utilizes the whole volume of input data, which can be regarded as a convolutional operation with kernel size being equal to the input volume in convolutional neural networks (CNN) [*Goodfellow et al.*, 2016; *LeCun et al.*, 1989]. The cost function is the mean squared errors between $\tau^d_{lk,NN}$ from DNN models and $\tau^d_{lk,DNS}$ from DNS. A validation dataset (20% of the total training data) is separated from the training dataset to detect overfitting. The input data is normalized by its standard deviation and mean in the whole DNS field. Each dense layer has a nonlinear Rectified Linear Unit (ReLU) [*Nair and Hinton*, 2010] activation function $f(x) = \max(0, x)$, which typically follows affine transformations in each layer [*Goodfellow et al.*, 2016]. Dropout [*Hinton et al.*, 2012] is used to avoid overfitting and to limit the influence of large weights. A schematic of the DNN set-up is shown in Figure S2. A mini-batch gradient descent method [*Ioffe and Szegedy*, 2015] is used to update the weight and to introduce some stochasticity in the descent by using small batches. The batch size is 1000 and number of epochs is 50.

| DNS data | $Re_\tau$ | $\frac{z_i}{L}$ | $\frac{L_x}{L_y}$ | $\frac{L_x}{L_z}$ | $\Delta_x^+$ | $\Delta_y^+$ | $\Delta_z^+$ |
|---|---|---|---|---|---|---|---|
| Sh2 | 309 | -678.2 | 1.5 | 6 | 2.87 | 2.87 | 0.71 |
| Sh5 | 554 | -105.1 | 1.5 | 6 | 4.95 | 4.95 | 1.19 |
| Sh20 | 1243 | -7.14 | 1.5 | 6 | 11.02 | 11.02 | 2.65 |
| Channel1 | 546 | 0 | 2 | $4\pi$ | 8.92 | 4.46 | 2.83 |

**Table S1.** *Details of DNS set-up in this study.* $Re_\tau$ is Reynolds number, $z_i$ is the boundary layer height, $L$ is the Obukhov length, $L_x$, $L_y$ and $L_z$ are the domain sizes in the $x$, $y$ and $z$ directions, respectively. $\Delta_x^+ = \frac{\Delta_x u_\tau}{\nu}$, $\Delta_y^+$ and $\Delta_z^+$ are the spatial grid resolutions denoted by inner units in the $x$, $y$ and $z$ directions, respectively. $u_\tau$ is the friction velocity and $\nu$ is the kinematic viscosity.

| Correlation for $\tau_{13}$ log-law region (whole DNS field) | M_uvw _multiSh | M_uv w | M_uv $w\theta$ | M_uv wp | M_uvw _multiT | M_uvw _multiC | M_uvw_ multiSh_ box555 | M_uvw_ multiSh_ box777 |
|---|---|---|---|---|---|---|---|---|
| $Sh20_{t=4,\Delta_z^+=11}$ | 0.847 (0.834) | 0.849 (0.837) | 0.847 (0.829) | 0.850 (0.851) | 0.850 (0.829) | 0.871 (0.884) | 0.890 (0.862) | 0.882 (0.853) |
| $Sh20_{t=4,\Delta_z^+=21}$ | 0.806 (0.840) | 0.804 (0.838) | 0.812 (0.843) | 0.809 (0.842) | 0.820 (0.850) | 0.750 (0.773) | 0.852 (0.876) | 0.858 (0.881) |
| $Sh20_{t=4,\Delta_z^+=42}$ | 0.716 (0.816) | 0.723 (0.817) | 0.738 (0.825) | 0.730 (0.822) | 0.744 (0.829) | 0.707 (0.803) | 0.765 (0.847) | 0.768 (0.851) |
| $Sh20_{t=4,\Delta_z^+=64}$ | 0.608 (0.778) | 0.611 (0.781) | 0.639 (0.798) | 0.636 (0.791) | 0.642 (0.797) | 0.476 (0.703) | 0.673 (0.808) | 0.669 (0.810) |
| $Sh2_{t=4,\Delta_z^+=6}$ | 0.793 (0.824) | 0.732 (0.769) | 0.753 (0.777) | 0.731 (0.763) | 0.788 (0.814) | 0.730 (0.757) | 0.838 (0.862) | 0.850 (0.873) |
| $Sh5_{t=1,\Delta_z^+=10}$ | 0.780 (0.818) | 0.732 (0.775) | 0.756 (0.787) | 0.734 (0.771) | 0.745 (0.787) | 0.718 (0.758) | 0.830 (0.858) | 0.840 (0.867) |

**Table S2.** *Correlation coefficients of $\tau_{13,DNS}$ and $\tau_{13,NN}$ predicted by different DNN models for different prediction datasets in both the log-law region and the whole DNS field.* Correlation coefficients for the whole DNS field are in the brackets. $\Delta_z^+ \equiv \Delta_z u_\tau/\nu$ is the spatial grid resolution in the vertical direction in inner units.

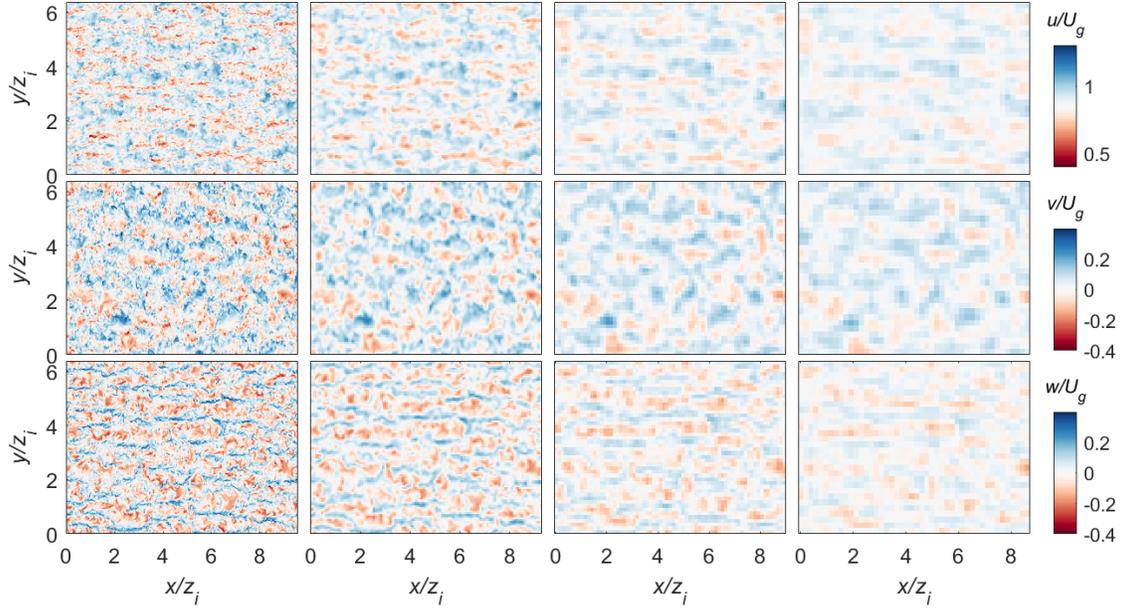

**Figure S1.** Velocity $u/U_g$ (first row), $v/U_g$ (second row) and $w/U_g$ (third row) of $x$-$y$ plane at $z^+ = \frac{z}{(\nu/u_\tau)} = 129.8$ in the log-law region of dataset Sh20. $U_g$ is geostrophic wind, $\nu$ is kinematic viscosity, $u_\tau$ is friction velocity and $z_i$ is boundary layer height. From left to right the selected datasets are $Sh20_{t=1,\Delta_z^+=3}$, $Sh20_{t=1,\Delta_z^+=21}$, $Sh20_{t=1,\Delta_z^+=42}$ and $Sh20_{t=1,\Delta_z^+=64}$.

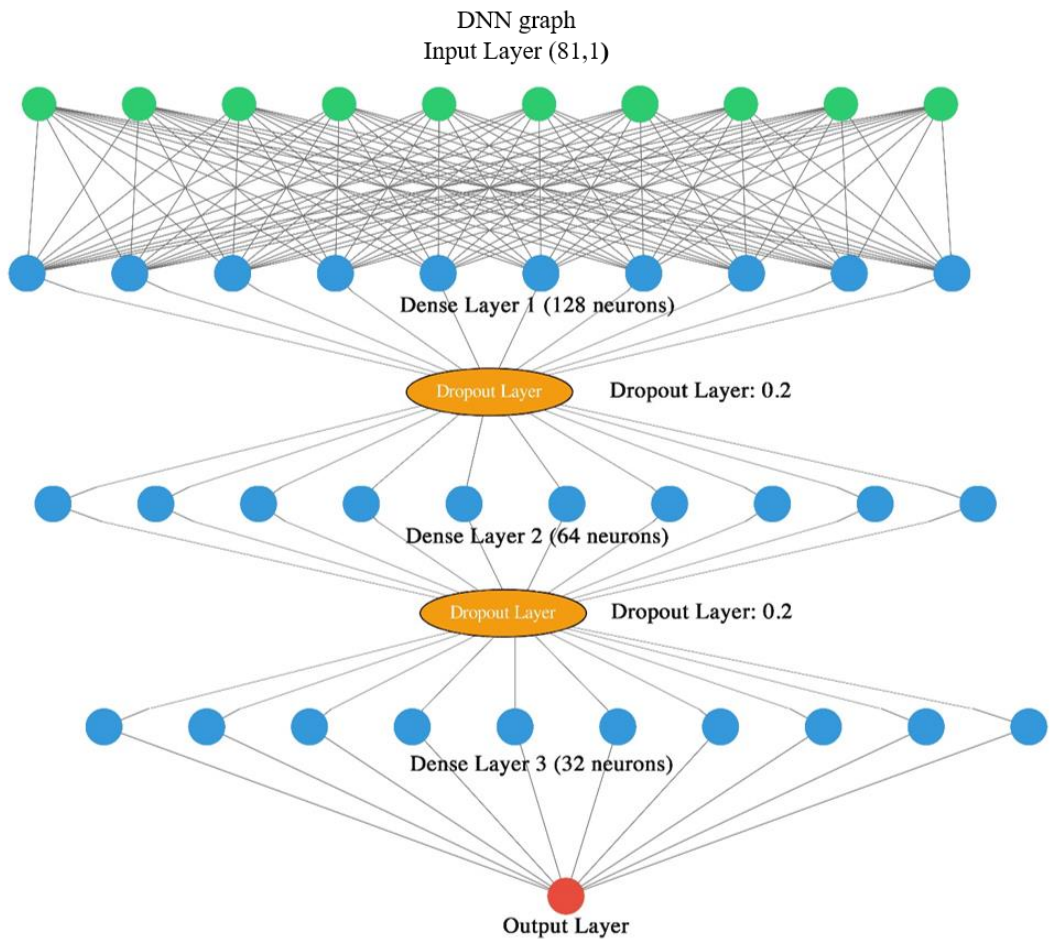

**Figure S2.** Schematic of the deep neural networks in this study.

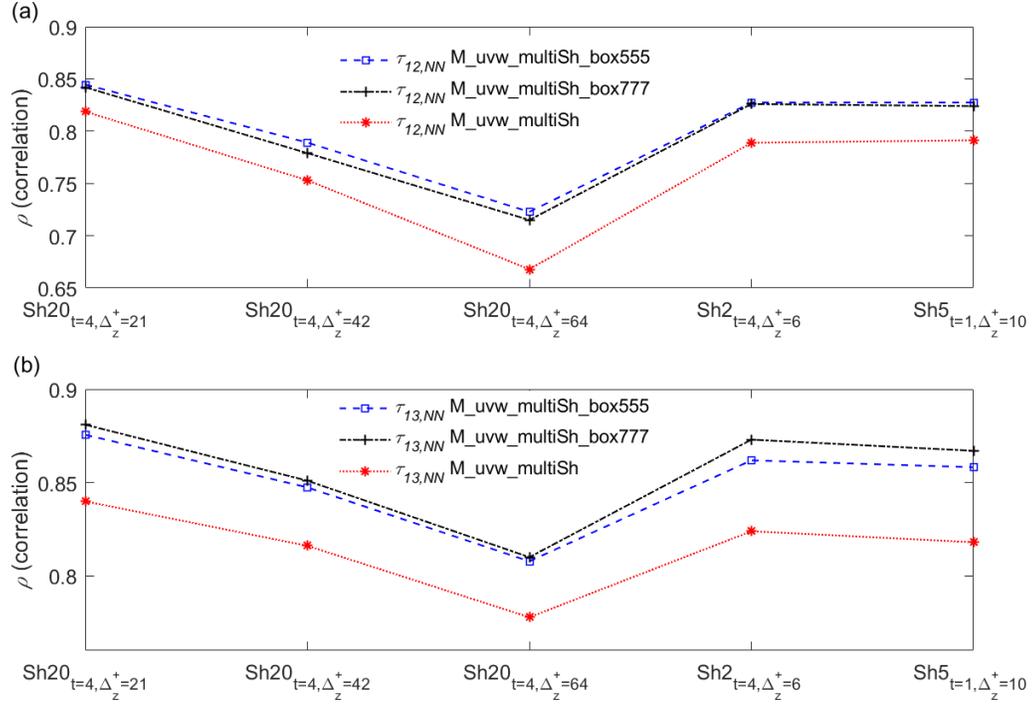

**Figure S3.** (a) Correlation coefficients between $\tau_{12,DNS}$ and $\tau_{12,NN}$ predicted from different DNN models in the whole DNS field. (b) Correlation coefficients between $\tau_{13,DNS}$ and $\tau_{13,NN}$ predicted from different DNN models in the whole DNS field. Details of the DNN models M_uvw_multiSh, M_uvw_multiSh_box555 and M_uvw_multiSh_box777 is shown in Table 1.

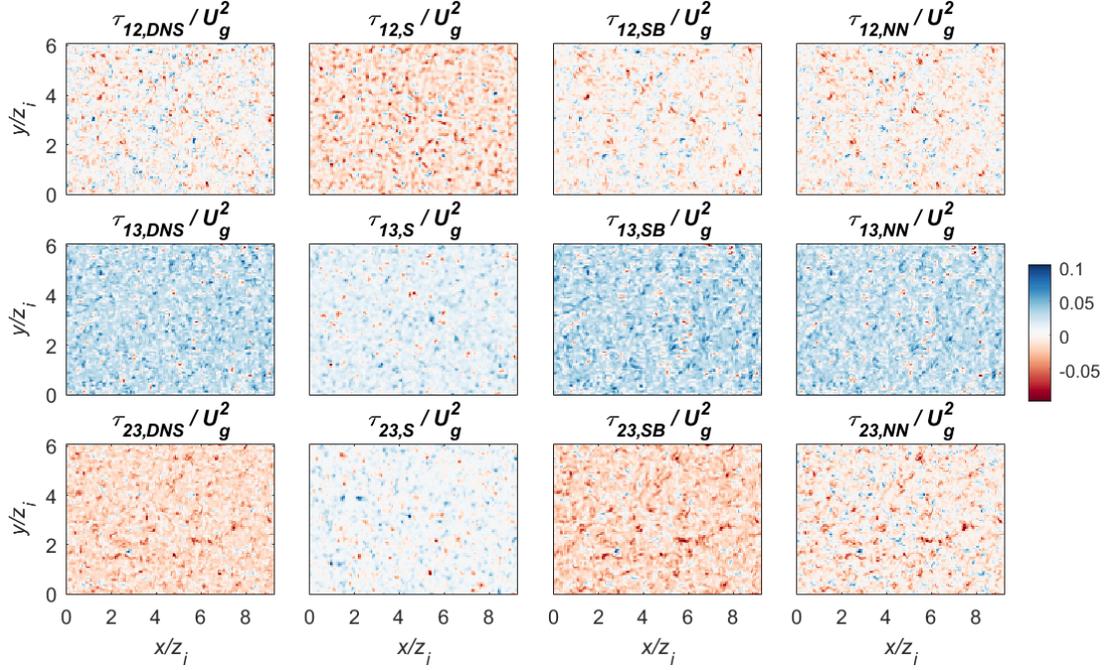

**Figure S4.** SGS stresses of $x$-$y$ plane normalized by $U_g^2$ at $z^+ = \frac{z}{(\nu/u_\tau)} = 77.3$ in the log-law region of dataset $Sh5_{t=1,\Delta_z^+=10}$. $U_g$ is geostrophic wind, $z_i$ is boundary layer height, $\nu$ is kinetic viscosity and $u_\tau$ is friction velocity. From left to right: SGS stresses calculated from DNS data, the Smagorinsky model, the Smagorinsky-Bardina mixed model and the DNN model M_uvw_multiSh.

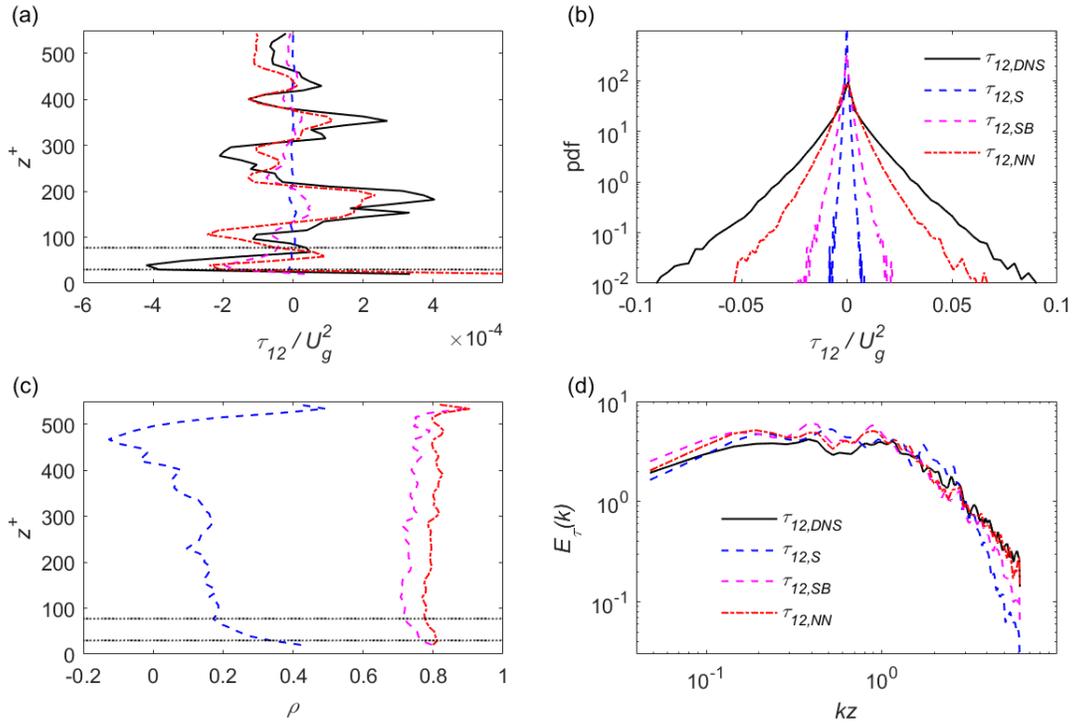

**Figure S5.** Same as figure 1 but for $\tau_{12}$.

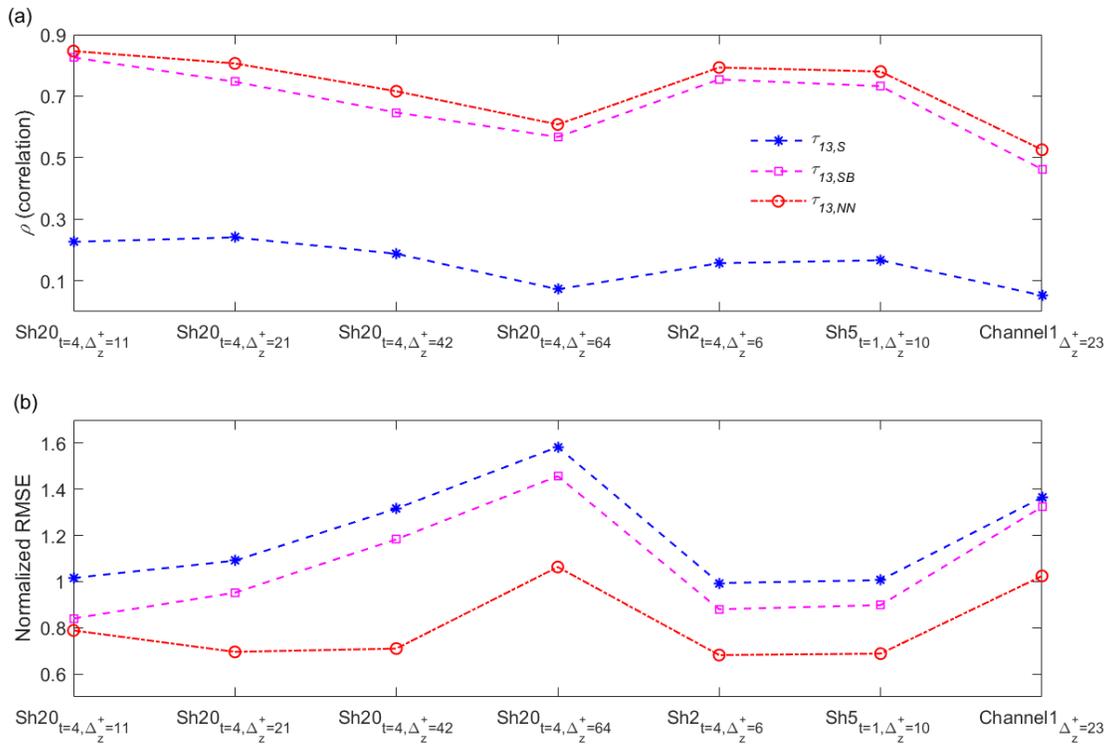

**Figure S6.** Same as figure 2 but only for the log-law region.

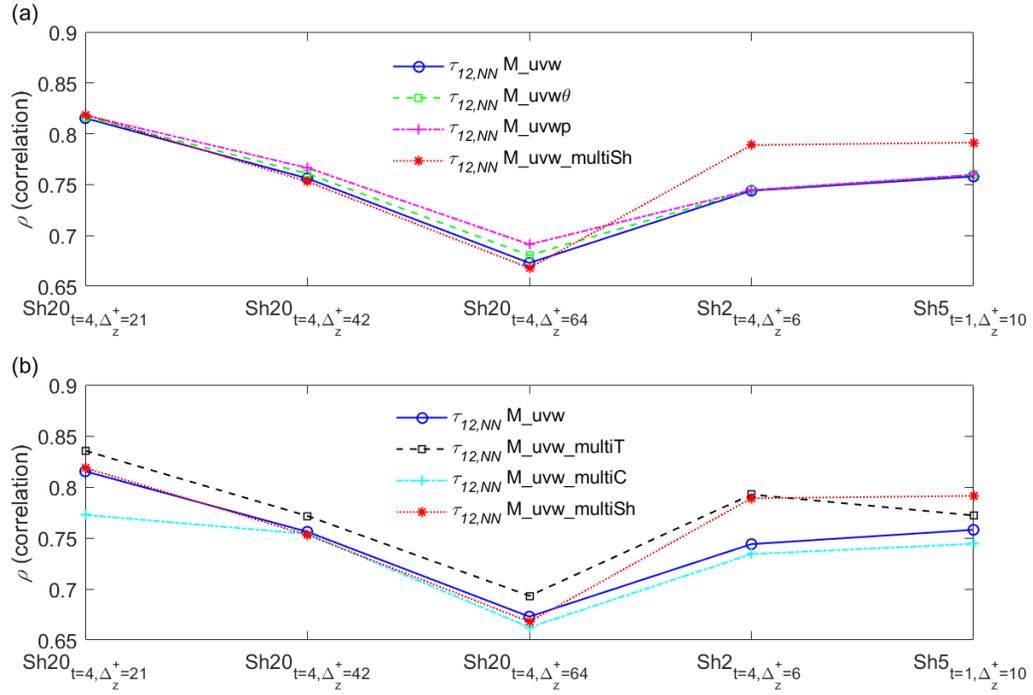

**Figure S7**. Same as Figure 3 but for $\tau_{12}$.

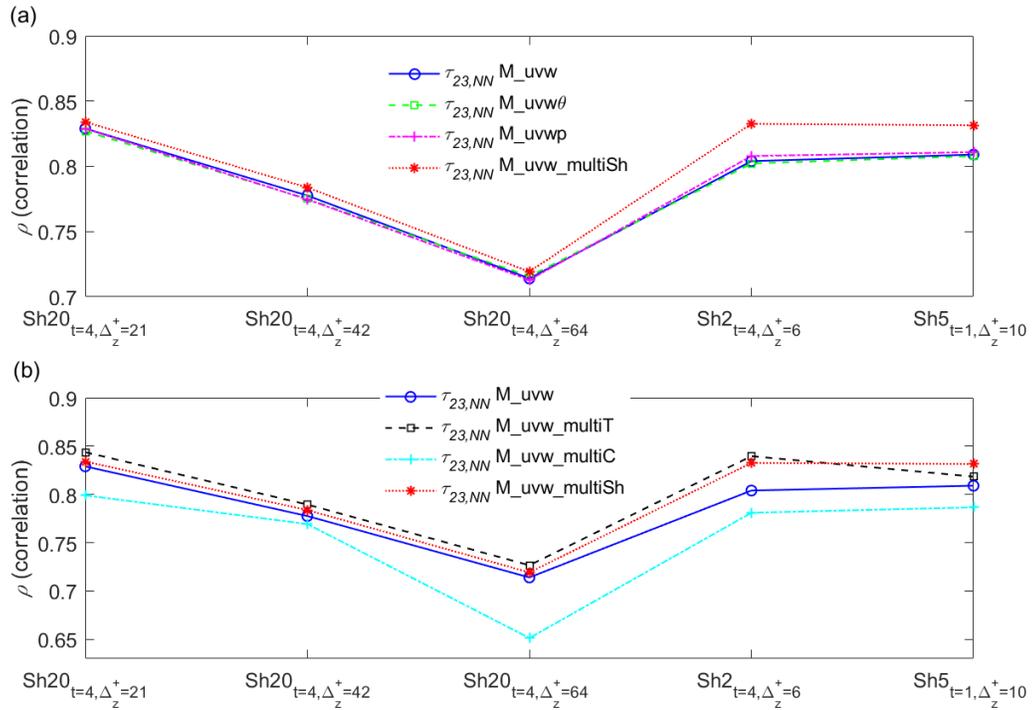

**Figure S8**. Same as Figure 3 but for $\tau_{23}$.

**References**

Abadi, M., P. Barham, J. Chen, Z. Chen, A. Davis, J. Dean, M. Devin, S. Ghemawat, G. Irving, and M. Isard (2016), Tensorflow: A system for large-scale machine learning, paper presented at 12th {USENIX} Symposium on Operating Systems Design and Implementation ({OSDI} 16)

Agee, E. M., T. Chen, and K. Dowell (1973), A review of mesoscale cellular convection, Bulletin of the American Meteorological Society, 54(10), 1004-1012. https://doi.org/10.1175/1520-0477(1973)054<1004:AROMCC>2.0.CO;2

Atkinson, B., and J. Zhang (1996), Mesoscale shallow convection in the atmosphere, Reviews of Geophysics, 34(4), 403-431. https://doi.org/10.1029/96RG02623

Bar-Sinai, Y., S. Hoyer, J. Hickey, and M. P. Brenner (2018), Data-driven discretization: a method for systematic coarse graining of partial differential equations, arXiv preprint arXiv:1808.04930

Bardina, J., J. Ferziger, and W. Reynolds (1980), Improved subgrid-scale models for large-eddy simulation, paper presented at 13th Fluid and PlasmaDynamics Conference. https://doi.org/10.2514/6.1980-1357

Bernard, P. S., and R. A. Handler (1990), Reynolds stress and the physics of turbulent momentum transport, Journal of Fluid Mechanics, 220, 99-124. https://doi.org/10.1017/S0022112090003202

Bou-Zeid, E., C. Higgins, H. Huwald, C. Meneveau, and M. B. Parlange (2010), Field study of the dynamics and modelling of subgrid-scale turbulence in a stable atmospheric surface layer over a glacier, Journal of Fluid Mechanics, 665, 480-515. https://doi.org/10.1017/S0022112010004015

Bradley, E. F., R. Antonia, and A. Chambers (1981), Turbulence Reynolds number and the turbulent kinetic energy balance in the atmospheric surface layer, Boundary-Layer Meteorology, 21(2), 183-197. https://doi.org/10.1007/BF02033936

Canuto, V., and Y. Cheng (1997), Determination of the Smagorinsky–Lilly constant CS, Physics of Fluids, 9(5), 1368-1378. https://doi.org/10.1063/1.869251

Chamecki, M., C. Meneveau, and M. B. Parlange (2007), The local structure of atmospheric turbulence and its effect on the Smagorinsky model for large eddy simulation, Journal of the Atmospheric Sciences, 64(6), 1941-1958. https://doi.org/10.1175/JAS3930.1

Cheng, Y., Q. Li, S. Argentini, C. Sayde, and P. Gentine (2018), Turbulence Spectra in the Stable Atmospheric Boundary Layer, arXiv preprint arXiv:1801.05847

Chollet, F., and Others (2015), Keras, https://keras.io

Chorin, A. J. (1968), Numerical solution of the Navier-Stokes equations, Mathematics of Computation, 22(104), 745-762. https://doi.org/10.1090/S0025-5718-1968-0242392-2

Chung, D., and G. Matheou (2014), Large-eddy simulation of stratified turbulence. Part I: A vortex-based subgrid-scale model, Journal of the Atmospheric Sciences, 71(5), 1863-1879. https://doi.org/10.1175/JAS-D-13-0126.1

Clark, R. A., J. H. Ferziger, and W. C. Reynolds (1979), Evaluation of subgrid-scale models using an accurately simulated turbulent flow, Journal of Fluid Mechanics, 91(1), 1-16. https://doi.org/10.1017/S002211207900001X

Corrsin, S. (1975), Limitations of gradient transport models in random walks and in turbulence, in Advances in Geophysics, edited by F. N. Frenkiel and R. E. Munn, 18, pp. 25-60, Elsevier. https://doi.org/10.1016/S0065-2687(08)60451-3

Deardorff, J. W. (1970), A numerical study of three-dimensional turbulent channel flow at large Reynolds numbers, Journal of Fluid Mechanics, 41(2), 453-480. https://doi.org/10.1017/S0022112070000691

Deardorff, J. W. (1972), Numerical investigation of neutral and unstable planetary boundary layers, Journal of the Atmospheric Sciences, 29(1), 91-115. https://doi.org/10.1175/1520-0469(1972)029<0091:NIONAU>2.0.CO;2

Domaradzki, J. A., W. Liu, and M. E. Brachet (1993), An analysis of subgrid-scale interactions in numerically simulated isotropic turbulence, Physics of Fluids A: Fluid Dynamics, 5(7), 1747-1759. https://doi.org/10.1063/1.858850

Dougherty, J. (1961), The anisotropy of turbulence at the meteor level, Journal of Atmospheric and Terrestrial Physics, 21(2-3), 210-213. https://doi.org/10.1016/0021-9169(61)90116-7

Egolf, P. W. (1994), Difference-quotient turbulence model: a generalization of Prandtl's mixing-length theory, Physical Review E, 49(2), 1260. https://doi.org/10.1103/PhysRevE.49.1260

Gamahara, M., and Y. Hattori (2017), Searching for turbulence models by artificial neural network, Physical Review Fluids, 2(5), 054604. https://doi.org/10.1103/PhysRevFluids.2.054604

Germano, M., U. Piomelli, P. Moin, and W. H. Cabot (1991), A dynamic subgrid-scale eddy viscosity model, Physics of Fluids A: Fluid Dynamics, 3(7), 1760-1765. https://doi.org/10.1063/1.857955

Giometto, M., A. Lozano-Durán, G. Park, and P. Moin (2017), Three-dimensional transient channel flow at moderate Reynolds numbers: Analysis and wall modeling*Rep.*, 193-205 pp, CTR Annu. Res. Briefs

Girshick, R., J. Donahue, T. Darrell, and J. Malik (2014), Rich feature hierarchies for accurate object detection and semantic segmentation, paper presented at Proceedings of the IEEE conference on computer vision and pattern recognition. https://doi.org/10.1109/CVPR.2014.81

Goodfellow, I., Y. Bengio, A. Courville, and Y. Bengio (2016), Deep Learning, MIT press Cambridge

Grachev, A. A., E. L. Andreas, C. W. Fairall, P. S. Guest, and P. O. G. Persson (2015), Similarity theory based on the Dougherty-Ozmidov length scale, Quarterly Journal of the Royal Meteorological Society, 141(690), 1845-1856. https://doi.org/10.1002/qj.2488

Härtel, C., L. Kleiser, F. Unger, and R. Friedrich (1994), Subgrid-scale energy transfer in the near-wall region of turbulent flows, Physics of Fluids, 6(9), 3130-3143. https://doi.org/10.1063/1.868137

Hinton, G. E., N. Srivastava, A. Krizhevsky, I. Sutskever, and R. R. Salakhutdinov (2012), Improving neural networks by preventing co-adaptation of feature detectors, arXiv preprint arXiv:1207.0580

Horiuti, K. (1989), The role of the Bardina model in large eddy simulation of turbulent channel flow, Physics of Fluids A: Fluid Dynamics, 1(2), 426-428. https://doi.org/10.1063/1.857465

Ioffe, S., and C. Szegedy (2015), Batch normalization: Accelerating deep network training by reducing internal covariate shift, arXiv preprint arXiv:1502.03167

Kleissl, J., C. Meneveau, and M. B. Parlange (2003), On the magnitude and variability of subgrid-scale eddy-diffusion coefficients in the atmospheric surface layer, Journal of the Atmospheric Sciences, 60(19), 2372-2388. https://doi.org/10.1175/1520-0469(2003)060<2372:OTMAVO>2.0.CO;2

Kleissl, J., M. B. Parlange, and C. Meneveau (2004), Field experimental study of dynamic Smagorinsky models in the atmospheric surface layer, Journal of the Atmospheric Sciences, 61(18), 2296-2307. https://doi.org/10.1175/1520-0469(2004)061<2296:FESODS>2.0.CO;2

Kolmogorov, A. N. (1941), The local structure of turbulence in incompressible viscous fluid for very large Reynolds numbers, paper presented at Dokl. Akad. Nauk SSSR

Krizhevsky, A., I. Sutskever, and G. E. Hinton (2012), Imagenet classification with deep convolutional neural networks, paper presented at Advances in Neural Information Processing Systems

LeCun, Y., B. Boser, J. S. Denker, D. Henderson, R. E. Howard, W. Hubbard, and L. D. Jackel (1989), Backpropagation applied to handwritten zip code recognition, Neural Computation, 1(4), 541-551. https://doi.org/10.1162/neco.1989.1.4.541

Leonard, A. (1975), Energy cascade in large-eddy simulations of turbulent fluid flows, in Advances in Geophysics, edited by F. N. Frenkiel and R. E. Munn, 18, pp. 237-248, Elsevier. https://doi.org/10.1016/S0065-2687(08)60464-1

Li, D., and E. Bou-Zeid (2011), Coherent structures and the dissimilarity of turbulent transport of momentum and scalars in the unstable atmospheric surface layer, Boundary-Layer Meteorology, 140(2), 243-262. https://doi.org/10.1007/s10546-011-9613-5

Li, Q., P. Gentine, J. P. Mellado, and K. A. McColl (2018), Implications of nonlocal transport and conditionally averaged statistics on Monin–Obukhov similarity theory and Townsend's attached eddy hypothesis, Journal of the Atmospheric Sciences, 75(10), 3403-3431. https://doi.org/10.1175/JAS-D-17-0301.1

Lilly, D. (1967), The representation of small-scale turbulence in numerical simulation experiments, paper presented at Proc. IBM Sci. Compt. Symp. Environ. Sci., White Plains, IBM, 1967

Lilly, D. K. (1992), A proposed modification of the Germano subgrid-scale closure method, Physics of Fluids A: Fluid Dynamics, 4(3), 633-635. https://doi.org/10.1063/1.858280

Ling, J., A. Kurzawski, and J. Templeton (2016), Reynolds averaged turbulence modelling using deep neural networks with embedded invariance, Journal of Fluid Mechanics, 807, 155-166. https://doi.org/10.1017/jfm.2016.615

Liu, S., C. Meneveau, and J. Katz (1994), On the properties of similarity subgrid-scale models as deduced from measurements in a turbulent jet, Journal of Fluid Mechanics, 275, 83-119. https://doi.org/10.1017/S0022112094002296

Lumley, J. L. (1981), Coherent structures in turbulence, in Transition and turbulence, edited by R. E. Meyer, pp. 215-242, Elsevier. https://doi.org/10.1016/B978-0-12-493240-1.50017-X

Marusic, I., J. P. Monty, M. Hultmark, and A. J. Smits (2013), On the logarithmic region in wall turbulence, Journal of Fluid Mechanics, 716(R3). https://doi.org/10.1017/jfm.2012.511

Mason, P. J. (1989), Large-eddy simulation of the convective atmospheric boundary layer, Journal of the Atmospheric Sciences, 46(11), 1492-1516. https://doi.org/10.1175/1520-0469(1989)046<1492:LESOTC>2.0.CO;2

Mason, P. J., and A. R. Brown (1999), On subgrid models and filter operations in large eddy simulations, Journal of the Atmospheric Sciences, 56(13), 2101-2114. https://doi.org/10.1175/1520-0469(1999)056<2101:OSMAFO>2.0.CO;2

McCulloch, W. S., and W. Pitts (1943), A logical calculus of the ideas immanent in nervous activity, The Bulletin of Mathematical Biophysics, 5(4), 115-133. https://doi.org/10.1007/BF02478259

McMillan, J., and J. H. Ferziger (1979), Direct testing of subgrid-scale models, AIAA Journal, 17(12), 1340-1346. https://doi.org/10.2514/3.61313

Meneveau, C., and J. Katz (2000), Scale-invariance and turbulence models for large-eddy simulation, Annual Review of Fluid Mechanics, 32(1), 1-32. https://doi.org/10.1146/annurev.fluid.32.1.1

Moeng, C.-H. (1984), A large-eddy-simulation model for the study of planetary boundary-layer turbulence, Journal of the Atmospheric Sciences, 41(13), 2052-2062. https://doi.org/10.1175/1520-0469(1984)041<2052:ALESMF>2.0.CO;2

Monin, A., and A. Obukhov (1954), Basic laws of turbulent mixing in the surface layer of the atmosphere, Contrib. Geophys. Inst. Acad. Sci. USSR, 151(163), e187

Nair, V., and G. E. Hinton (2010), Rectified linear units improve restricted boltzmann machines, paper presented at Proceedings of the 27th International Conference on Machine Learning (ICML-10)

Nicoud, F., F. J. F. Ducros, turbulence, and Combustion (1999), Subgrid-scale stress modelling based on the square of the velocity gradient tensor, Flow, Turbulence and Combustion, 62(3), 183-200. https://doi.org/10.1023/A:1009995426001

Nieuwstadt, F. T., P. J. Mason, C.-H. Moeng, and U. Schumann (1993), Large-eddy simulation of the convective boundary layer: A comparison of four computer codes, in Turbulent Shear Flows 8, edited by F. Durst, R. Friedrich, B. E. Launder, F. W. Schmidt, U. Schumann and J. H. Whitelaw, pp. 343-367, Springer, Berlin, Heidelberg. https://doi.org/10.1007/978-3-642-77674-8_24

Obukhov, A. (1946), Turbulence in thermally inhomogeneous atmosphere, Trudy Inst. Teor. Geofiz. Akad. Nauk SSSR, 1, 95-115

Orlandi, P. (2012), Fluid flow phenomena: a numerical toolkit, Springer Science & Business Media. https://doi.org/10.1007/978-94-011-4281-6

Ozmidov, R. (1965), On the turbulent exchange in a stably stratified ocean. Izv. Acad. Sci. USSR, Atmos. Oceanic Phys., 1, 861-871

Panofsky, H. A. (1963), Determination of stress from wind and temperature measurements, Quarterly Journal of the Royal Meteorological Society, 89(379), 85-94. https://doi.org/10.1002/qj.49708937906

Paulson, C. A. (1970), The mathematical representation of wind speed and temperature profiles in the unstable atmospheric surface layer, Journal of Applied Meteorology, 9(6), 857-861. https://doi.org/10.1175/1520-0450(1970)009<0857:TMROWS>2.0.CO;2

Pope, S. (2000), Turbulent flows, Cambridge University Press, Cambridge. https://doi.org/10.1017/CBO9780511840531

Porté-Agel, F., M. B. Parlange, C. Meneveau, and W. E. Eichinger (2001), A priori field study of the subgrid-scale heat fluxes and dissipation in the atmospheric surface layer, Journal of the Atmospheric Sciences, 58(18), 2673-2698. https://doi.org/10.1175/1520-0469(2001)058<2673:APFSOT>2.0.CO;2

Salesky, S. T., M. Chamecki, and E. Bou-Zeid (2017), On the nature of the transition between roll and cellular organization in the convective boundary layer, Boundary-Layer Meteorology, 163(1), 41-68. https://doi.org/10.1007/s10546-016-0220-3

Schmitt, F. G. (2007), About Boussinesq's turbulent viscosity hypothesis: historical remarks and a direct evaluation of its validity, Comptes Rendus Mécanique, 335(9-10), 617-627. https://doi.org/10.1016/j.crme.2007.08.004

Sermanet, P., D. Eigen, X. Zhang, M. Mathieu, R. Fergus, and Y. LeCun (2013), Overfeat: Integrated recognition, localization and detection using convolutional networks, arXiv preprint arXiv:1312.6229

Shah, S. K., and E. Bou-Zeid (2014), Direct numerical simulations of turbulent Ekman layers with increasing static stability: modifications to the bulk structure and second-order statistics, Journal of Fluid Mechanics, 760, 494-539. https://doi.org/10.1017/jfm.2014.597

Smagorinsky, J. (1963), General circulation experiments with the primitive equations: I. The basic experiment, Monthly Weather Review, 91(3), 99-164. https://doi.org/10.1175/1520-0493(1963)091<0099:GCEWTP>2.3.CO;2

Speziale, C. G. (1991), Analytical methods for the development of Reynolds-stress closures in turbulence, Annual Review of Fluid Mechanics, 23(1), 107-157. https://doi.org/10.1146/annurev.fl.23.010191.000543

Taigman, Y., M. Yang, M. A. Ranzato, and L. Wolf (2014), Deepface: Closing the gap to human-level performance in face verification, paper presented at Proceedings of the IEEE conference on computer vision and pattern recognition. https://doi.org/10.1109/CVPR.2014.220

Tennekes, H., J. L. Lumley, and J. Lumley (1972), A first course in turbulence, MIT press

Voller, V., and F. Porte-Agel (2002), Moore's law and numerical modeling, Journal of Computational Physics, 179(2), 698-703. https://doi.org/10.1006/jcph.2002.7083

von Kármán, T. (1930), Mechanische ahnlichkeit und turbulenz, paper presented at Proceedings of the 3rd International Congress on Applied Mechanics

Wray, A. A. (1990), Minimal storage time advancement schemes for spectral methods, NASA Ames Research Center, California, Report No. MS, 202

Xu, D., and J. Chen (2016), Subgrid-scale dynamics and model test in a turbulent stratified jet with coexistence of stable and unstable stratification, Journal of Turbulence, 17(5), 443-470. https://doi.org/10.1080/14685248.2015.1129407

Yang, X., S. Zafar, J.-X. Wang, and H. Xiao (2019), Predictive large-eddy-simulation wall modeling via physics-informed neural networks, Physical Review Fluids, 4(3), 034602. https://doi.org/10.1103/PhysRevFluids.4.034602